\documentclass[12pt]{article}
\usepackage{longtable,supertabular}
\usepackage{amsmath}
\usepackage{amssymb}
\usepackage{latexsym}
\usepackage{cite,a4wide,color}
\usepackage{multirow}
\usepackage{ntheorem}
\usepackage[square,numbers]{natbib}

\newtheorem{theorem}{Theorem}[section]
\newtheorem{corollary}{Corollary}[section]
\newtheorem{proposition}{Proposition}[section]
\newtheorem{lemma}{Lemma}[section]
\newtheorem{definition}{Definition}[section]
\newtheorem*{proof}{Proof.}

\newtheorem{example}{Example}[section]

\theoremstyle{remark}

\newcommand{\m}{\mathrm{m}}
\newcommand{\ord}{\mathrm{ord}}
\newcommand{\rank}{\mathrm{rank}}
\newcommand{\SR}{\mathrm{SR}}

\newcommand{\bC}{{\mathbf{{C}}}}
\newcommand{\bF}{{\mathbf{{F}}}}
\newcommand{\bT}{{\mathbf{{T}}}}

\newcommand{\bc}{{\mathbf{c}}}

\newcommand{\lcm}{{\mathrm{lcm}}}

\title{\bf Cyclic and Negacyclic Sum-Rank Codes}
\author{Hao Chen, Cunsheng Ding, Zhiqiang Cheng and Conghui Xie\thanks{Hao Chen, Zhiqiang Cheng and Conghui Xie are with the College of Information Science and Technology/Cyber Security, Jinan University, Guangzhou, Guangdong Province, 510632, China (e-mail: haochen@jnu.edu.cn, zqcheng@stu2023.jnu.edu.cn, conghui@stu2021.jnu.edu.cn). C. Ding is with the Department of Computer Science and Engineering, The Hong Kong University of Science and Technology, Hong Kong, China (e-mail: cding@ust.hk).
		The research of Hao Chen was supported by NSFC Grant 62032009. The research of Cunsheng Ding was supported by the Hong Kong Research Grants Council under Grant No. 16301123.
}}

\begin{document}
	
	\maketitle
	\begin{abstract}
		Sum-rank codes have known applications in the multishot network coding, the distributed storage and the construction of space-time codes. U. Mart\'{\i}nez-Pe\~{n}as introduced
	the	cyclic-skew-cyclic sum-rank codes and proposed the BCH bound on the cyclic-skew-cyclic sum-rank codes in his paper published in IEEE Trans. Inf. Theory, vol. 67, no.  8, 2021. Afterwards, many sum-rank BCH codes with lower bounds on their dimensions and minimum sum-rank distances were constructed.   Sum-rank Hartmann-Tzeng bound and sum-rank Roos bound on cyclic-skew-cyclic codes were proposed and proved by G. N. Alfarano, F. J. Lobillo, A. Neri, and A. Wachter-Zeh in 2022. In this paper,  cyclic,  negacyclic and constacyclic sum-rank codes
		are introduced and a direct construction of cyclic, negacyclic and constacyclic sum-rank codes of the matrix size $m \times m$ from cyclic,  negacyclic and constacyclic codes over ${\bf F}_{q^m}$ in the Hamming metric is proposed.
		The cyclic-skew-cylic sum-rank codes are special cyclic sum-rank codes.
In addition,   BCH and Hartmann-Tzeng bounds for a type of cyclic sum-rank codes are developed.
Specific constructions of cyclic, negacyclic and constacyclic sum-rank codes with known dimensions and controllable
minimum sum-rank distances are proposed.\\

Moreover,  many distance-optimal binary sum-rank codes and an infinite family of distance-optimal binary cyclic sum-rank codes with minimum sum-rank distance four are constructed. This is the first infinite family of distance-optimal sum-rank codes with minimum sum-rank distance four in the literature.\\
		
		Constacyclic sum-rank code.  Distance-optimal sum-rank codes.
	\end{abstract}
	
	\newpage
	\section{Introduction}
	
	The Hamming weight $wt_H({\bf a})$ of a vector ${\bf a}=(a_1, \ldots, a_n) \in {\bf F}_q^n$ is the cardinality  of its support $$supp({\bf a})=\{i:a_i \neq 0\}.$$ The Hamming distance $d_H({\bf a}, {\bf b})$ between two vectors ${\bf a}$ and ${\bf b}$ is defined to be the Hamming weight of ${\bf a}-{\bf b}$. For a code ${\bf C} \subset {\bf F}_q^n$, its minimum Hamming distance is $$d_H=\min_{{\bf a} \neq {\bf b}} \{d_H({\bf a}, {\bf b}): {\bf a}, {\bf b} \in {\bf C}\}.$$ It is well-known that the Hamming distance of a linear code ${\bf C}$ is the minimum Hamming weight of its non-zero codewords. For the theory of error-correcting codes in the Hamming metric,  the reader is referred to \cite{MScode,Lint,HP}.   A code ${\bf C}$ in ${\bf F}_q^n$ of  minimum distance $d_H$ and cardinality $M$ is called an $(n, M, d_H)_q$ code. The code rate is $R({\bf C})=\frac{\log_q M}{n},$ and the relative distance is $\delta({\bf C})=\frac{d_H}{n}.$ For an $[n, k, d_H]_q$ linear code, the Singleton bound asserts that $d_H \leq n-k+1$. When the equality holds, this code is called a maximal distance separable (MDS) code.  Reed-Solomon codes are well-known MDS codes \cite{HP}.
Some BCH codes and Goppa codes can be considered as subfield subcodes of generalized Reed-Solomon codes, and contain examples of optimal binary codes for many parameters (\cite{MScode,HP,codetable}).\\

	The rank-metric distance on the space ${\bf F}_q^{(n, m)}$ of size $n \times m$ matrices over ${\bf F}_q$, where $n \leq m$, is defined by  $d_r(A,B)= \rank(A-B)$. The minimum rank-distance of a code ${\bf C} \subset {\bf F}_q^{(n,m)}$  is $$d_r({\bf C})=\min_{A\neq B} \{d_r(A,B): A, B \in {\bf C} \}.$$  For a code ${\bf C}$ in ${\bf F}_q^{(m, n)}$ with the minimum rank distance $d_r({\bf C}) \geq d$, the Singleton-like bound asserts that the number of codewords in ${\bf C}$ is upper bounded by $q^{m(n-d+1)}$  \cite{Gabidulin}.  A code satisfying the equality is called a maximal rank distance (MRD) code. The Gabidulin code in ${\bf F}_q^{(n, n)}$ consists of ${\bf F}_q$-linear mappings on ${\bf F}_q^n \cong {\bf F}_{q^n}$ defined by $q$-polynomials $a_0x+a_1x^q+\cdots+a_ix^{q^i}+\cdots+a_tx^{q^t}$, where $a_t,\ldots,a_0$ are arbitrary elements in ${\bf F}_{q^n}$ \cite{Gabidulin}. Then the minimum rank-distance of the Gabidulin code is at least $n-t$ since each nonzero $q$-polynomial above has at most $q^t$ roots in ${\bf F}_{q^n}$. There are $q^{n(t+1)}$ such $q$-polynomials. Hence the size of the Gabidulin code is $q^{n(t+1)}$ and it is an MRD code. \\
	
	Sum-rank codes were introduced in \cite{NU} for their applications in multishot network coding (see also  \cite{MK19,NPS}).
	The sum-rank codes were used to construct space-time codes \cite{SK}, and codes for distributed storage (\cite{CMST,MK,MP1}). There have been some papers on upper bounds, constructions and applications of sum-rank codes in recent years (see, e.g., \cite{BGR1,BGR,MP1,MP191,MK19,MP20,MP21,Neri,NSZ21,MP22} and references therein).
	Fundamental properties and some bounds on sizes of sum-rank codes were given in \cite{BGR}. For a nice survey of sum-rank codes and their applications, the reader is referred to \cite{MPK22}.\\

	We recall some basic concepts and results for sum-rank codes from \cite{BGR}. Let ${\bf F}_q^{(n,m)}$ be the set of all $ n \times m$ matrices over $\bF_q$. This is a linear space over ${\bf F}_q$ of dimension $nm$. Let $n_i \leq m_i$ be $2t$ positive integers satisfying $m_1 \geq m_2 \geq \cdots \geq m_t$. Set $N=n_1+\cdots+n_t$.  Let
$${\bf F}_q^{(n_1, m_1), \ldots,(n_t, m_t)}=
{\bf F}_q^{(n_1,  m_1)} \bigoplus \cdots \bigoplus {\bf F}_q^{(n_t, m_t)}$$
be the set of all ${\bf x}=({\bf x}_1,\ldots,{\bf x}_t)$, ${\bf x}_i \in {\bf F}_q^{(n_i \times m_i)}$, $i=1,\ldots,t$. This is a linear space over ${\bf F}_q$ of dimension $\Sigma_{i=1}^t n_im_i$.
Set $$wt_{sr}({\bf x}) =\Sigma_{i=1}^t \rank({\bf x}_i),$$
where ${\bf x}=({\bf x}_1, \ldots, {\bf x}_t) \in {\bf F}_q^{(n_1, m_1), \ldots,(n_t, m_t)}$. The sum-rank distance between two vectors ${\bf x}$ and ${\bf y}$ in ${\bf F}_q^{(n_1, m_1), \ldots,(n_t, m_t)}$ is $$d_{sr}({\bf x},{\bf y})=wt_{sr}({\bf x}-{\bf y})=\Sigma_{i=1}^t \rank({\bf x}_i-{\bf y}_i),$$
where ${\bf x}=({\bf x}_1, \ldots, {\bf x}_t)$ and ${\bf y}=({\bf y}_1, \ldots, {\bf y}_t)$ in ${\bf F}_q^{(n_1,m_1), \ldots,(n_t,m_t)}$. This is indeed a metric on ${\bf F}_q^{(n_1,m_1), \ldots,(n_t,m_t)}$. \\
	
A subset $\bC$ of  the finite space ${\bf F}_q^{(n_1,m_1), \ldots,(n_t,m_t)}$ with the sum-rank metric is called
a \emph{sum-rank code} with {\it block length} $t$ and {\it matrix sizes} $n_1 \times m_1, \ldots, n_t\times m_t$.
Its minimum sum-rank distance is defined by $$d_{sr}({\bf C})=\min_{{\bf x} \neq {\bf y}, {\bf x}, {\bf y} \in {\bf C}} d_{sr}({\bf x}, {\bf y}).$$  The code rate of ${\bf C}$ is $R_{sr}=\frac{\log_q |{\bf C}|}{\Sigma_{i=1}^t n_im_i}$. The relative distance is $\delta_{sr}=\frac{d_{sr}}{N}$. When ${\bf C} \subset {\bf F}_q^{(n_1,m_1), \ldots,(n_t,m_t)}$ is a linear subspace of the linear space ${\bf F}_q^{(n_1,m_1), \ldots,(n_t,m_t)}$ over ${\bf F}_q$, ${\bf C}$ is called a {\it linear sum-rank code}. When $q=2$, this code is called a {\it binary sum-rank code}. \\

	In general it is interesting to construct good sum-rank codes with large sizes and large minimum sum-rank distances, and develop their decoding algorithms. When $t=1$, this is the rank-metric code case  \cite{Gabidulin}.  When $m_1=\cdots=m_t=m$ and $n_1=\cdots=n_t=n$, this is the $t$-sum-rank code over ${\bf F}_{q^m}$ with length $nt$.  When $m=n=1$, this is the Hamming metric code case. Hence the sum-rank coding is a generalization of the the Hamming metric coding and the rank-metric coding. When $q=2$ and $n_1=\cdots=n_t=m_1=\cdots=m_t=2$, sum-rank codes are in the space ${\bf F}_2^{(2,2), \ldots, (2,2)}={\bf F}_2^{(2, 2)} \bigoplus \cdots \bigoplus {\bf F}_2^{(2, 2)}$, this is the generalization of the binary Hamming metric codes to the sum-rank codes of the matrix size $2 \times 2$. In the fundamental paper \cite{MP21} by U. Mart\'{\i}nez-Pe\~{n}as, many linear binary sum-rank BCH codes of matrix size $2\times 2$ were constructed, and their decoding algorithms were given.\\

	The Singleton-like bound on sum-rank codes was proposed in \cite{MK,BGR}. The general form of
	the bound given in Theorem III.2 in \cite{BGR} is as follows. The minimum sum-rank distance $d_{sr}$ can be written uniquely in the form $d_{sr}=\Sigma_{i=1}^{j-1} n_i+\delta+1$ where $0 \leq \delta \leq n_j-1$,
	then $$|{\bf C}| \leq q^{\Sigma_{i=j}^t n_im_i-m_j\delta}.$$ A code attaining this bound is called a
	{\it maximal sum-rank distance (MSRD) code.} When $m_1=\cdots=m_t=m$, this bound is of the form $$|{\bf C}| \leq q^{m(N-d_{sr}+1)}.$$  It is clear that when $t=1$, this is the Singleton-like bound for rank-metric codes, and when $n=m=1$, this is the Singleton bound for codes in the Hamming metric. For constructions of some MSRD codes,  the reader is referred to \cite{BGR,MP20,Chen}. Linearized Reed-Solomon codes, which are the sum-rank versions of the classical Reed-Solomon codes, were introduced and studied in \cite{MP1}. Their Welch-Berlekamp decoding algorithm was presented in \cite{MK19}.\\

	There have been several constructions of sum-rank codes with good parameters.  The analogue sum-rank codes of Hamming codes and simplex codes were given in \cite{MP191}. Sum-rank Hamming codes have the minimum sum-rank distance three and are perfect. The reader is referred to \cite{MP1}
and \cite{MP21} for linearized Reed-Solomon codes. Twisted linearized Reed-Solomon codes were constructed in \cite{Neri}. Cyclic-skew-cyclic (CSC) sum-rank codes of the matrix size $n_1=\cdots=n_t=n$, $m_1=\cdots=m_t=m$ were proposed and studied in \cite{MP21} by a deep algebraic method. These sum-rank codes are linear over ${\bf F}_{q^m}$.  The sum-rank BCH bound on CSC codes was proved and sum-rank BCH codes were introduced in \cite{MP21}. Dimensions of sum-rank BCH codes were lower bounded in \cite[Theorem 9]{MP21}. Many sum-rank codes of the parameters $n=m=2$ and $q=2$ were constructed in Tables I, II, III, IV, V, VI and VII of \cite{MP21}. The minimum sum-rank distances of CSC codes were improved in a recent paper \cite{ALNWZ}. Similar sum-rank Hartmann-Tzeng bound and sum-rank Roos bound for CSC codes were proved in their paper. A decoding algorithm for sum-rank BCH codes was presented in \cite{MP21}. For several recent works on sum-rank codes,  the reader is referred to \cite{AKR,Berardini,Borello,Chen2}.
In our previous paper \cite{Chen},  we proposed a construction of linear sum-rank codes by combining several Hamming metric codes and $q$-polynomials.  \\

Similar to the distance-optimal codes in the Hamming metric, see \cite{Ding5}, distance-optimal sum-rank codes can be defined as follows. Let $\bC \subset {\bf F}_q^{(n_1,m_1), \ldots,(n_t,m_t)}$ be a sum-rank code with $M$ codewords and the minimum sum-rank distance $d_{sr}$, and there is no sum-rank code with $M$ codewords and the minimum sum-rank distance $d_{sr}+1$, we call $\bC$ a distance-optimal sum-rank code. We will construct an infinite family of distance-optimal cyclic sum-rank codes with respect to the sphere packing bound. We refer to \cite{BGR} for the sphere packing bound in the sum-rank metric. In general, it is more difficult to construct optimal sum-rank codes attaining some upper bounds in the sum-rank metric. Sum-rank Hamming codes constructed in \cite{MP191} are perfect in the sum-rank metric, then distance-optimal. However the minimum sum-rank distance of sum-rank Hamming codes is three. In this paper, we show that cyclic sum-rank codes provide infinitely many distance-optimal sum-rank codes with minimum sum-rank distance four.\\

	The main contributions of this paper are as follows.\\

1) Cyclic,  negacyclic and constacyclic  sum-rank codes over ${\bf F}_q$ of the matrix size $n\times m$ are introduced. Cyclic-skew-cyclic sum-rank codes are special cyclic sum-rank codes.\\

2) Cyclic,  negacyclic  and constacyclic sum-rank codes are constructed with cyclic, negacyclic  and constacyclic codes in the Hamming metric through a one-way bridge.  \\

3) With cyclic or negacyclic or  constacyclic codes in the Hamming metric with known
dimensions,  specific families of cyclic or negacyclic or constacyclic  sum-rank codes with known dimensions
and controllable minimum sum-rank distances are presented.\\

4) In some cases,  the minimum sum-rank distances of  some cyclic, negacyclic and constacyclic sum-rank codes are determined explicitly.  Notice that
it is difficult to determine the minimum sum-rank distances and dimensions of cyclic-skew-cyclic sum-rank codes (\cite{MP21,ALNWZ}).\\

5) Many distance-optimal binary sum-rank codes and an infinite family of distance-optimal binary cyclic sum-rank codes with minimum sum-rank distance four are constructed. To the best of the authors' knowledge, this is the first infinite family of distance-optimal codes with minimum sum-rank distance four in the literature.\\

The rest of this paper is organized as follows.
In Section \ref{sec-CSC}, cyclic-skew-cyclic sum-rank codes and their bounds are recalled.
In Section \ref{sec-2}, cyclic and negacyclic sum-rank codes are defined, and a general construction of cyclic and negacyclic sum-rank codes is introduced. In addition, the BCH bound and Hartmann-Tzeng bound for a type of cyclic sum-rank codes are developed. In Section \ref{sec-3}, specific families of cyclic sum-rank codes are presented.
 In Section \ref{sec-4}, constructions of negacyclic sum-rank codes are proposed.
 In Section \ref{sec-n1}, $\lambda$-constacyclic sum-rank codes are introduced and studied.
 In Section \ref{sec-chenh}, distance-optimal binary sum-rank codes with minimum sum-rank distance four are constructed.
 In Section \ref{sec-5},  the decoding of cyclic and negacyclic sum-rank codes is discussed.
Section \ref{sec-6} concludes this paper.

\section{Cyclic-skew-cyclic sum-rank codes and their bounds}\label{sec-CSC}	

We first recall the definition of cyclic-skew-cyclic (CSC) sum-rank codes introduced and studied in \cite{MP21,ALNWZ}.  By Definition 4 in \cite{MP21},  a sum-rank code ${\bf C}$ over ${\bf F}_q$
with block length $t$ and matrix size $n \times m$ is called {\it cyclic-skew-cyclic (CSC)} if
the following hold:

(1) If ${\bf x}=({\bf x}_1,\ldots,{\bf x}_t)\in {\bf C}$ , where ${\bf x}_i \in {\bf F}_q^{(n,m)}$,
$i=1,\ldots,t$, then $({\bf x}_{t},{\bf x}_1,...,{\bf x}_{t-1})\in {\bf C}$.

(2)  If ${\bf x}=({\bf x}_1,\ldots,{\bf x}_t)\in {\bf C}$ with ${\bf x}_i = (x_{i,1},x_{i,2},...,x_{i,m})
\in {\bf F}_q^{(n,m)}$, $i=1,\ldots,t$,
then $({\bf x}'_{1},{\bf x}'_2,...,{\bf x}'_{t})\in {\bf C}$, where ${\bf x}'_i =(\sigma(x_{i,m}),\sigma(x_{i,1}),...,\sigma(x_{i,m-1}))\in {\bf F}_q^{(n,m)}$, $i=1,\ldots,t$,
and $\sigma: {\bf F}_{q^{ns}}\rightarrow {\bf F}_{q^{ns}}$ is defined as $a \longmapsto a^{q^s}$ with $s=\ord_t(q)$.\\

In \cite{MP21} and \cite{ALNWZ}, sum-rank BCH lower bound, sum-rank Hartmann-Tzeng lower bound and sum-rank Roos lower bound on minimum sum-rank distances of CSC codes were given.  We recall these bounds for references later. \\
	
We consider sum-rank codes over $\bF_q$  with block length $t$ and matrix size $n \times m$.
Assume that $\gcd(t, q)=1$.
Let
$$x^t-1=m_1(x) \cdots m_h(x)$$
be the canonical factorisation of $x^t-1$ over $\bF_q$, where each $m_i(x)$  is an irreducible polynomial in ${\bf F}_q[x]$.
Let $s=\ord_t(q)$. Let $a\in {\bf F}_{q^s}$ be a primitive $t$-th root of unity and $\beta$ be a normal element of the extension ${\bf F}_{q^{s}} \subseteq {\bf F}_{q^{ns}}$. Given a cyclic-skew-cyclic code ${\bf C} \subseteq {\bf F}_q^{(n,m),...,(n,m)}$ with minimal generator skew polynomial $g(x,z)$, the defining set of ${\bf C}$ is
${\bf T_C}=\{(a,\beta)\in {\bf F}_{q^{s}}\times {\bf F}_{q^{ns}}^*|a^t=1,$ total evaluation map Ev$_{a,\beta}(g(x,z))=0\}.$
	
	\begin{theorem}[Sum-rank BCH bound, \cite{MP21}]\label{NT-2-4}
		If the defining set ${\bf T_C}$ of the cyclic-skew-cyclic code ${\bf C}$ is the smallest that contains $\{(a^{b+i},\sigma^{i}(\beta)) \in {\bf F}_{q^s}\times {\bf F}_{q^{ns}}^*:0\leq i\leq \delta-2\}$, where $b\geq 0$ and $2\leq\delta\leq mt$,
		then $d_{sr}\geq\delta$,  $\dim_{{\bf F}_{q}}({\bf C})\geq n(mt-\sum\limits_{\mu=1}^{h}d_\mu\cdot \min\{n,\frac{sk_\mu}{d_\mu}\}) $,
		where $d_\mu=\deg_x(m_\mu(x))$ and $k_\mu=| \{i|0\leq i \leq \delta-2,m_\mu(a^{b+i})=0\} |$.
	\end{theorem}
	
	The sum-rank HT bound and sum-rank Roos bound were proposed and proved in \cite{ALNWZ}. Lower bounds on dimensions of cyclic-skew-cyclic sum-rank codes can be obtained from these two bounds as follows.
	\begin{theorem}[Sum-rank HT bound, \cite{ALNWZ}]\label{T-2-5}
		If the defining set ${\bf T_C}$ of the cyclic-skew-cyclic code ${\bf C}$ is the smallest that contains $\{(a^{b+it_1+jt_2},\sigma^{it_1+jt_2}(\beta)) \in {\bf F}_{q^s}\times {\bf F}_{q^{ns}}^*:0\leq i\leq \delta-2, 0\leq j \leq r\}$, where $\gcd(mt,t_1)=1$ and $\gcd(mt,t_2)<\delta$,
		then $d_{sr}\geq\delta+r$, $\dim_{{\bf F}_{q}}({\bf C})\geq n(mt-\sum\limits_{\mu=1}^{h}d_\mu\cdot \min\{n,\frac{sk_\mu}{d_\mu}\}) $,
		where $d_\mu=\deg_x(m_\mu(x))$ and $k_\mu=| \{(b+it_1+jt_2)|0\leq i\leq \delta-2, 0\leq j \leq r,m_\mu(a^{b+it_1+jt_2})=0\} |$.
	\end{theorem}	
	
	\begin{theorem}[Sum-rank Roos bound, \cite{ALNWZ}]\label{T-2-6}
		Let ${\bf C}$ be a cyclic-skew-cyclic code. Let $b$,$u$ $\delta$, $k_0$,...,$k_r$ be integers, such that $\gcd(mt, u) =1$, $k_i < k_{i+1}$ for $i=0,...,r-1$, $k_r-k_0\leq\delta+r-1$. If $\{(a^{b+ui+k_j},\sigma^{ui+k_j}(\beta)) \in {\bf F}_{q^s}\times {\bf F}_{q^{ns}}^*:0\leq i\leq \delta-2, 0\leq u \leq r\}\subseteq {\bf T}_{\bf C}$, 	
		then $d_{sr}({\bf C}) \geq \delta + r$,  $\dim_{{\bf F}_{q}}({\bf C})\geq n(mt-\sum\limits_{\mu=1}^{h}d_\mu\cdot \min\{n,\frac{sk_\mu}{d_\mu}\}) $, where $d_\mu=\deg_x(m_\mu(x))$ and $k_\mu=| \{(b+ui+k_j)|0\leq i\leq \delta-2, 0\leq u \leq r,m_\mu(a^{b+ui+k_j})=0\} |.$
	\end{theorem}

When $n=m=1$, above lower bounds are the classical BCH bound, the classical Hartmann-Tzeng bound and the classical Roos bound for cyclic codes in the Hamming metric.\\

	\section{Cyclic and negacyclic sum-rank codes}\label{sec-2}
	
	In this section, we first introduce cyclic and negacyclic sum-rank codes in ${\bf F}_q^{(n, m), \ldots,(n, m)}$. Cyclic-skew-cyclic(CSC) codes introduced in \cite{MP21} are special cases of cyclic sum-rank codes.
	Then we  present a general construction of cyclic sum-rank codes directly from cyclic codes in the Hamming metric.
Afterwards, we propose and prove the BCH bound and Hartmann-Tzeng bound for a type of cyclic sum-rank codes.

\subsection{The classical constacyclic codes in the Hamming metric}

We recall some basic facts of cyclic, negacyclic and constacyclic codes in the Hamming matric.
Let $\lambda$ be a nonzero element in ${\mathbf F}_q.$ A linear code ${\bf C}\subseteq  {\mathbf F}_{q}^n$ over ${\mathbf F}_{q}$ is said to be {\it $\lambda$-constacyclic} if $(c_0,c_1,...,c_{n-1})\in {\bf C}$ implies $(\lambda c_{n-1},c_0,c_1,...,c_{n-2})\in {\bf C}$. By identifying any vector $(c_0,c_1,...,c_{n-1})\in {\mathbf F}_{q}^n$ with $$c_0+c_1x+c_2x^2+\cdots+c_{n-1}x^{n-1}\in {\mathbf F}_{q}[x]/(x^n-\lambda),$$ the corresponding constacyclic code ${\bf C}$ of length $n$ over ${\mathbf F}_{q}$ is an ideal of ${\mathbf F}_{q}[x]/(x^n-\lambda)$. Note that every ideal of ${\mathbf F}_{q}[x]/(x^n-\lambda)$ is principal. Then there exists monic polynomial $g(x)$ of the smallest degree such that ${\bf C}=\left \langle g(x) \right \rangle$ and $g(x)|(x^n-\lambda)$. In addition, $g(x)$ is unique and called the {\it generator polynomial}.  The polynomial $h(x)=(x^n-\lambda)/g(x)$ is called
the \emph{check polynomial} of the constacyclic code $\bC$.
In particular, if $\lambda=1$ or $\lambda=-1$,  the $\lambda$-constacyclic code is {\it cyclic} or {\it negacyclic},  respectively (\cite{HP,MScode}). \\

When a code in ${\bf F}_q^n$ is only closed with respect to the addition, not the scalar multiplication, and is cyclic or negacyclic, we call this code a {\it cyclic additive code} or a {\it negacyclic cyclic code}.
Cyclic additive codes were introduced and studied in \cite{Bier}.

\subsection{Lower bounds on cyclic codes in the Hamming metric}
	
Assume that $\gcd(n,q)=1.$	Let $s=\ord_n(q)$ and $\alpha$ be a generator of the cyclic group ${\mathbf F}_{{q}^s}^*$ and put $\beta=\alpha^{\frac{{q}^s-1}{n}}$. Then $\beta$ is a primitive $n$-th root of unity.
Let $\bC$ be a cyclic code of length $n$ over $\bF_q$ with generator polynomial $g(x)$.
The {\it defining set} of ${\bf C}$ with respect to $\beta$ is defined by
${\bf T}=\{0\leq i\leq n-1|g(\beta^i)=0\}$.  \\
	
If there are an integer $b$ and an integer $\delta \geq 2$ such that
$$\{(b+i) \bmod{n}: 0 \leq i \leq \delta -2 \} \subseteq  {\bf T},$$
then ${\bf T}$ is said to contain $\delta-1$ consecutive elements and the minimum distance $d_H(\bC)$ of ${\bf C}$ is at least $\delta$ \cite{HP}.
This is the {\it BCH bound} for cyclic codes.  \\

Let $A$ be a set of $\delta-1$ consecutive elements of $\bT$ and $B=\{jb \bmod{n}| 0\leq j\leq s\}$,
where $\gcd(b,n)< \delta$. If $A+B\subseteq \bT$, then the minimum distance $d_H(\bC)$ of ${\bf C}$
is at least $\delta+s$. This bound is called the {\it Hartmann-Tzeng (HT) bound} for cyclic codes.
For more bounds on cyclic codes,  the reader is referred to \cite{HP,MScode}.\\

\subsection{Cyclic and negacyclic sum-rank codes}

Cyclic-skew-cyclic sum-rank codes were introduced in Section \ref{sec-CSC}.  We
now define cyclic sum-rank codes as follows.

	\begin{definition}[Cyclic sum-rank code]
		A sum-rank code ${\bf C}$ over ${\bf F}_q$ with block length $t$ and matrix size $n \times m$ is called a {\it cyclic sum-rank (CSR) code}, if ${\bf x}=({\bf x}_1,\ldots,{\bf x}_t)$ is a codeword in ${\bf C}$,
		${\bf x}_i \in {\bf F}_q^{(n,m)}$, $i=1,\ldots,t$,  implies that
		$({\bf x}_{t},{\bf x}_1,...,{\bf x}_{t-1})$ is also a codeword in ${\bf C}$.
	\end{definition}

Similarly, negacyclic sum-rank codes are defined as follows.

\begin{definition}[Negacyclic sum-rank code]
		A sum-rank code ${\bf C}$ over ${\bf F}_q$ with block length $t$ and matrix size $n \times m$ is called a {\it negacyclic sum-rank code}, if ${\bf x}=({\bf x}_1,\ldots,{\bf x}_t)$ is a codeword in ${\bf C}$,
		${\bf x}_i \in {\bf F}_q^{(n,m)}$, $i=1,\ldots,t$,  implies that
		$({\bf -x}_{t},{\bf x}_1,...,{\bf x}_{t-1})$ is also a codeword in ${\bf C}$.
\end{definition}

The linear space ${\bf F}_q^{(m,m)}$ of all $m \times m$ matrices over ${\bf F}_q$ can be identified with the space of all $q$-polynomials, $${\bf F}_q^{(m,m)}=\{a_0x+a_1x^q+\cdots+a_{m-1}x^{q^{m-1}}: a_0, \ldots, a_{m-1} \in {\bf F}_{q^m}\}.$$ This isomorphism is the isomorphism of the linear spaces over ${\bf F}_q$. Then each sum-rank code ${\bf C}$ of the matrix size $m \times m$ corresponds $m$ codes ${\bf C}_0, \ldots, {\bf C}_{m-1}$ over ${\bf F}_{q^m}$ via this isomorphism. When ${\bf C}$ is a cyclic or negacyclic sum-rank code, ${\bf C}_0, \ldots, {\bf C}_{m-1}$ are cyclic additive or negacyclic additive codes over ${\bf F}_{q^m}$. The closeness with respect to the scalar multiplication over ${\bf F}_{q^m}$ is not necessarily satisfied.  \\

Now we are ready to introduce a method for constructing linear sum-rank codes over $\bF_q$
with block length $t$ and matrix size $m \times m$ using a sequence of $m$ linear codes
$\bC_0, \ldots, \bC_{m-1}$ over $\bF_{q^m}$ with length $t$.
Given linear codes ${\bf C}_i\subseteq {\bf F}_{q^{m}}^t$ with parameters  $[t,k_i,d_i]_{q^{m}}$,
$0 \leq i \leq m-1$,
the sum-rank code $\SR({\bf C}_0,...,{\bf C}_{m-1})$ is
defined by
\begin{eqnarray}
\SR({\bf C}_0,...,{\bf C}_{m-1})=
\left\{   \SR(\bc_0, \ldots, \bc_{m-1}):
\bc_i=(c_{i,1}, \ldots, c_{i,t}) \in \bC_i, \ 0 \leq i \leq m-1
\right\},
\end{eqnarray}
where
\begin{eqnarray}
 \SR(\bc_0, \ldots, \bc_{m-1}):=
\left(\sum_{i=0}^{m-1} c_{i,1} x^{q^i}, \ldots,      \sum_{i=0}^{m-1} c_{i,t} x^{q^i} \right).
\end{eqnarray}
We sometimes use ${\bf c}_0x+{\bf c}_1x^q+\cdots+{\bf c}_{m-1}x^{q^{m-1}}$  to denote the codeword
$\SR({\bf c}_0, \ldots, {\bf c}_{m-1})$.
 $\SR({\bf C}_0,...,{\bf C}_{m-1})$  is a linear sum-rank code over $\bF_q$.
 It is easy to see that
 $$\dim_{{\bf F}_q}(\SR({\bf C}_0,...,{\bf C}_{m-1}))=m(k_0+\cdots +k_{m-1}).$$
The following is an improved version of Theorem 2.1 in \cite{Chen}.

\begin{theorem}\label{thm-mainmain}
	Let ${\bf C}_i \subset {\bf F}_{q^m}^t$ be a $[t, k_i, d_i]_{q^m}$ linear code over ${\bf F}_{q^m}$, $0 \leq i \leq m-1$. Then $\SR({\bf C}_0,...,{\bf C}_{m-1})$ is a  linear sum-rank code  over ${\bf F}_q$ with block length $t$, matrix size $m \times m$ and dimension  $m(k_0+\cdots +k_{m-1})$.
	The minimum sum-rank distance of $\SR({\bf C}_0,{\bf C}_{1},...,{\bf C}_{m-1})$ is at least
	$$\max\{\min \{md_0, (m-1)d_1,...,d_{m-1}\}, \min\{d_0,2d_1, \ldots, md_{m-1}\}\}.$$
\end{theorem}

\noindent
{\bf Proof:}
Let $\SR(\bc_0, \ldots, \bc_{m-1}) \in \SR({\bf C}_0,...,{\bf C}_{m-1})$ be any nonzero codeword,  where
$\bc_i=(c_{i,1}, \ldots, c_{i,t}) \in {\bf C}_i$ for $0 \leq i \leq m-1$.  Then at least one of these $\bc_i$ is a nonzero codeword.
Suppose that ${\bf c}_j$ is a nonzero codeword of ${\bf C}_j$ but ${\bf c}_i$ is the zero codeword of
${\bf C}_i$ for all $j+1 \leq i \leq m-1$, where $0 \leq j \leq m-1$.
Then at each coordinate position $i$ of $\SR({\bf C}_0, \ldots, {\bf C}_{m-1})$, $1 \leq i \leq t$,  we have
\begin{eqnarray*}
&& \rank(c_{0,i}x+c_{1,i}x^q+\cdots+c_{m-1,i}x^{q^{m-1}}) \\
&& =\rank(c_{0,i}x+c_{1,i}x^q+\cdots+c_{j,i}x^{q^j}) \\
&& \geq (m-j),
\end{eqnarray*}
since $c_{0,i}x+c_{1,i}x^q+\cdots+c_{j,i}x^{q^j}$ has at most $q^j$ roots in ${\bf F}_{q^m}$.
Therefore, $$wt_{sr}(\SR(\bc_0, \ldots, \bc_{m-1}) ) \geq (m-j)d_j.$$
Consequently,  the minimum sum-rank distance of $\SR({\bf C}_0,...,{\bf C}_{m-1})$ is at least
$$\min \{md_0, (m-1)d_1,...,d_{m-1}\}. $$

On the other hand,  let ${\bf c}_i$ be the zero codeword of ${\bf C}_i$ for all $0 \leq i \leq j-1$
but ${\bf c}_j$ be a nonzero codeword of ${\bf C}_j$.  Then at the $i$-th coordinate position of
$\SR({\bf C}_0, \ldots, {\bf C}_{m-1})$,  we have
$$\rank(c_{0,i}x+c_{1,i}x^q+\cdots+c_{m-1,i}x^{q^{m-1}})=\rank(c_{j,i}x^{q^j}+\cdots+c_{m-1,i}x^{q^{m-1}}).$$ Since the ${\bf F}_q$-linear mapping $c_{j,i}x^{q^j}+\cdots+c_{m-1,i}x^{q^{m-1}}$
of ${\bf F}_{q^m}$ is the same as the ${\bf F}_q$-linear mapping $(c_{j,i}^{q^{m-j}}x+\cdots c_{m-1,i}^{q^{m-j}}x^{q^{m-1-j}})^{q^j}$, $c_{j,i}x^{q^j}+\cdots+c_{m-1,i}x^{q^{m-1}}$ and $c_{j,i}^{q^{m-j}}x+\cdots c_{m-1,i}^{q^{m-j}}x^{q^{m-1-j}}$ have the same rank. Then we have $$\rank(c_{j,i}q^j+\cdots+c_{m-1,i}x^{q^{m-1}}) \geq j+1.$$
Consequently,  the minimum sum-rank distance of $\SR({\bf C}_0,...,{\bf C}_{m-1})$ is at least
$$\min \{d_0, 2d_1, \ldots, md_{m-1}\}. $$
This completes the proof. \\

	The following result gives an upper bound on the minimum sum-rank distance of the sum-rank code $\SR({\bf C}_0,...,{\bf C}_{m-1})$.
	\begin{proposition}\label{P-3.1}
		Let ${\bf C}_i \subset {\bf F}_{q^m}^t$ be a $[t, k_i, d_i]_{q^m}$ linear code over ${\bf F}_{q^m}$, $0 \leq i \leq m-1$. Then $\SR({\bf C}_0,...,{\bf C}_{m-1})$ is a  linear sum-rank code  over ${\bf F}_q$ with block length $t$, matrix size $m \times m$ and dimension  $m(k_0+\cdots +k_{m-1})$.
		The minimum sum-rank distance of $\SR({\bf C}_0,{\bf C}_{1},...,{\bf C}_{m-1})$ is at most
		$$\min \{md_0, md_1,...,md_{m-1}\}.$$	
	\end{proposition}
	\noindent
	{\bf Proof:}
	Let $\bc_i=(c_{i,1}, \ldots, c_{i,t})\in {\bf C}_i$ and $wt_H({\bf c}_i)=d_i$ for $0\leq i\leq m-1.$ Then $$(c_{i,1}x^{q^{i}}, c_{i,2}x^{q^{i}},\cdots, c_{i,t}x^{q^{i}})\in \SR({\bf C}_0,...,{\bf C}_{m-1}).$$
	Assume that $c_{i,j} \neq 0$.  Since the mapping $c_{i,j}x^{q^{i}}$ has only one root zero in ${\bf F}_{q^m}$,
	we have
	 $$\rank(c_{i,j}x^{q^{i}})=m.$$
	Hence the minimum sum-rank distance of $\SR({\bf C}_0,{\bf C}_{1},...,{\bf C}_{m-1})$ is at most
	$$\min \{md_0, md_1,...,md_{m-1}\}.$$	
	This completes the proof. \\
	
	Combining Theorem \ref{thm-mainmain} and Proposition \ref{P-3.1} yields the following proposition,
	which shows that $d_{sr}(\SR({\bf C}_0,{\bf C}_{1},...,{\bf C}_{m-1}))$ can be determined explicitly in some cases.
	
	\begin{proposition}\label{P-3.2}
				Let ${\bf C}_i \subset {\bf F}_{q^m}^t$ be a $[t, k_i, d_i]_{q^m}$ linear code over ${\bf F}_{q^m}$, $0 \leq i \leq m-1$. Then $\SR({\bf C}_0,...,{\bf C}_{m-1})$ is a  linear sum-rank code  over ${\bf F}_q$ with block length $t$, matrix size $m \times m$ and dimension  $m(k_0+\cdots +k_{m-1})$. If $md_0\leq \min\{(m-1)d_1,(m-2)d_2,...,d_{m-1}\} ({\rm or}~ md_{m-1}\leq \min\{d_0,2d_1...,(m-1)d_{m-2}\}),$
		then the minimum sum-rank distance of $\SR({\bf C}_0,{\bf C}_{1},...,{\bf C}_{m-1})$ is exactly $md_0 ({\rm or}~md_{m-1}).$	
	\end{proposition}

We have the following important remarks on the construction of $\SR({\bf C}_0,...,{\bf C}_{m-1})$.
\begin{itemize}
\item When each ${\bf C}_i\subset {\bf F}_{q^m}^t$ is a cyclic or negacyclic  code for $0\leq i\leq m-1$,
it is clear that $\SR({\bf C}_0,...,{\bf C}_{m-1})$ is a cyclic or negacyclic sum-rank code.  This allows
us to construct cyclic (respectively,  negacyclic) sum-rank codes with cyclic (respectively, negacyclic)
codes in the Hamming metric.
\item Note that the locations of the coefficients in the $q$-polynomial $a_0x+a_1x^q + \cdots + a_{m-1}x^{q^{m-1}}$ over $\bF_{q^m}$ play different roles in determining the rank of the $\bF_q$-linear mapping.
The order of the using these linear codes $\bC_i$ in the construction of $\SR({\bf C}_0,...,{\bf C}_{m-1})$
affects the minimum sum-rank distance of $\SR({\bf C}_0,...,{\bf C}_{m-1})$.  However, it does not affect
the dimension of $\SR({\bf C}_0,...,{\bf C}_{m-1})$.
\item The lower bound on the minimum sum-rank distance of $\SR({\bf C}_0,...,{\bf C}_{m-1})$
is in general tight, as it can be achieved by certain codes (see Examples \ref{exam-4.1},  \ref{exam-4.2},
\ref{exam-4.3}, \ref{exam-4.4} and Proposition 3.2 and 4.1).
\item This construction $\SR({\bf C}_0,...,{\bf C}_{m-1})$ could produce very good linear sum-rank codes
(see Example \ref{exam-4.4}).
\item It is open if every linear sum-rank code over $\bF_q$ with block size $t$ and matrix size $m \times m$
can be constructed as $\SR({\bf C}_0,...,{\bf C}_{m-1})$ using a sequence of $m$ codes $\bC_i$ over
$\bF_{q^m}$ with length $t$.
\item It is open if every cyclic sum-rank code over $\bF_q$ with block size $t$ and matrix size $m \times m$
can be constructed as $\SR({\bf C}_0,...,{\bf C}_{m-1})$ using a sequence of $m$ cyclic codes $\bC_i$ over
$\bF_{q^m}$ with length $t$.
\end{itemize}
	
Since the first condition in the definition of cyclic-skew-cylic codes is the same as the condition for cyclic sum-rank codes, cyclic-skew-cyclic sum-rank codes are special cyclic sum-rank codes. However,
a cyclic sum-rank code is not necessarily  a cyclic-skew-cyclic code.
The cyclic sum-rank code below  is not a cyclic-skew-cyclic code \cite{MP21}.

\begin{example}
{\rm Let ${\mathbf C}_0 \subset \mathbf{F}_4^{15}$  be a cyclic code over $\mathbf{F}_4$ with generator polynomial $g_1(x)=x^{2}+x+{\alpha}$, where $\alpha$ is a primitive element of ${\bf F}_4$. Then $\alpha^2=\alpha+1$ and $\{1,\alpha\}$ is an ordered base of ${\bf F}_4$ over ${\bf F}_2$. Let ${\mathbf C}_1\subset \mathbf{F}_4^{15}$  be a cyclic code over $\mathbf{F}_4$ with generator polynomial $g_2(x)=\frac{x^{15}-1}{x-1}$.  Set $\SR({\bf C}_0,{\bf C}_1)=\{{\bf c}_0x+{\bf c}_1x^2 : {\bf c}_0=(c_{0,1},...,c_{0,15})\in {\mathbf C}_0, {\bf c}_1=(c_{1,1},...,c_{1,15})\in {\mathbf C}_1\}.$ Then $\SR({\bf C}_0,{\bf C}_1)$ is a cyclic sum-rank code. For $i=0,1$, let $a_i=a_{i1}+a_{i2}\alpha,$ where $a_{i1},a_{i2}\in {\bf F}_2,$
		then the linear map $a_{1}x+a_{2}x^2$ over ${\bf F}_4$ corresponds to matrix $\begin{bmatrix}
			a_{1,1}+a_{2,1} & a_{1,2}+a_{2,2}+a_{2,1} \\
			a_{1,2}+a_{2,2} & a_{1,1}+a_{2,1}+a_{1,2} \\
		\end{bmatrix}$ under the ordered base $\{1,\alpha\}$. Conversely,  the matrix $\begin{bmatrix}
			a & b \\
			c & d \\
		\end{bmatrix}$  of a linear map over ${\bf F}_4$ under the ordered base $\{1,\alpha\}$ corresponds to linear map $a'_{1}x+a'_{2}x^2$, where $a'_{1}=a+b+c+(a+d)\alpha$ and $a'_{2}=b+c+(a+c+d)\alpha$. \\

 Suppose that $\SR({\bf C}_0,{\bf C}_1)$ is a cyclic-skew-cyclic sum-rank code. From the condition $s=4$ and $q=2$, it follows that $\sigma(a)=a$ for $a \in {\bf F}_4$. Take
 $${\bf c}_0=  [1+\alpha,\alpha,\alpha,0,0,0,0,0,0,0,0,0,0,0,0]\in \mathbf {C}_0$$ and
 $${\bf c}_1=[1,1,1,1,1,1,1,1,1,1,1,1,1,1,1] \in \mathbf {C}_1, $$
 then ${\bf c}_0x+{\bf c}_1x^2$ corresponds to $$\SR({\bf c}_0,{\bf c}_1)=\left(
		\begin{bmatrix}
			0 & 0 \\
			1 & 1 \\
		\end{bmatrix}
		,
		\begin{bmatrix}
			1 & 0 \\
			1 & 0 \\
		\end{bmatrix}
		,
		\begin{bmatrix}
			1 & 0 \\
			1 & 0 \\
		\end{bmatrix}
		,
		\begin{bmatrix}
			1 & 1 \\
			0 & 1 \\
		\end{bmatrix},
		\begin{bmatrix}
			1 & 1 \\
			0 & 1 \\
		\end{bmatrix},...,			
		\begin{bmatrix}
			1 & 1 \\
			0 & 1 \\
		\end{bmatrix}
		\right).$$
		From the second condition on cyclic-skew-cyclic sum-rank codes, we change the first and second columns of it to
		$$\SR({\bf c}'_0,{\bf c}'_1)=\left(
		\begin{bmatrix}
			0 & 0 \\
			1 & 1 \\
		\end{bmatrix}
		,
		\begin{bmatrix}
			0 & 1 \\
			0 & 1 \\
		\end{bmatrix}
		,
		\begin{bmatrix}
			0 & 1 \\
			0 & 1 \\
		\end{bmatrix}
		,
		\begin{bmatrix}
			1 & 1 \\
			1 & 0 \\
		\end{bmatrix},
		\begin{bmatrix}
			1 & 1 \\
			1 & 0 \\
		\end{bmatrix},...,			
		\begin{bmatrix}
			1 & 1 \\
			1 & 0 \\
		\end{bmatrix}
		\right),$$
		where ${\bf c}'_0=(1+\alpha,1+\alpha,...,1+\alpha)$ and ${\bf c}'_1=(1,1+\alpha,1+\alpha,0,0,...,0).$
		Then ${\bf c}'_1=(1,1+\alpha,1+\alpha,0,0,...,0) \in {\bf C}_2$, a contradiction since every coordinate of the codeword of ${\bf C}_1$ is equal.
		Hence $\SR({\bf C}_0,{\bf C}_1)$ does not satisfy the second condition of cyclic-skew-cyclic sum-rank codes. $\SR({\bf C}_0,{\bf C}_1)$ is not a cyclic-skew-cyclic code.  It is only a cyclic sum-rank code.}
	
\end{example}

	\subsection{Cyclic sum-rank codes with the matrix size $m \times m$}

We have the following BCH bound and the Hartmann-Tzeng bound for cyclic sum-rank codes $\SR(\bC_0, \ldots, \bC_{m-1})$ with matrix size $m \times m$.

	\begin{theorem}[BCH bound for cyclic sum-rank codes $\SR(\bC_0, \ldots, \bC_{m-1})$]\label{T-2-8}
		Let ${\bf C}_i$ $ \subseteq {\bf F}_{q^m}^t$ be a cyclic code with defining set ${\bf T}_i$,
		$0 \leq i\leq m-1$.  If  ${\bf T}_i$ contains $\delta_i-1$ consecutive elements, $0\leq i\leq m-1$,
		 then the $q$-ary cyclic sum-rank code $\SR({\bf C}_0,...,{\bf C}_{m-1})$ has
		dimension $m(mt-\sum\limits_{i=0}^{m-1}|{\bf T}_i|)$ and minimum sum-rank distance at least
		  $$
		   \max\{ \min \{m\delta_0, (m-1)\delta_1,...,\delta_{m-1}\},
		  \min \{\delta_0, 2\delta_1,...,m\delta_{m-1}\}
		   \}
		  $$	
	\end{theorem}
	
	\begin{proof}
		{\rm The desired conclusions follow from Theorem \ref{thm-mainmain} and the BCH bound for cyclic codes in the Hamming metric.}
	\end{proof}

	\begin{theorem}[Hartmann-Tzeng bound for $\SR(\bC_0, \ldots, \bC_{m-1})$]\label{T-2-9}
		Let ${\bf C}_i$ $ \subseteq {\bf F}_{q^m}^t$ be a cyclic code with defining set ${\bf T}_i$,
		$0 \leq i\leq m-1$.   		
		If  $A_i$ is a set of $\delta_{i}-1$ consecutive elements in ${\bf T}_i$ and there exists
		$B_i=\{jb_i \bmod{t} |0\leq j\leq s_i\}$, where $\gcd(b_i,t)< \delta_i$,
		satisfying $A_i+B_i\subseteq {\bf T}_i$, $0 \leq i \leq m-1$,
		then  $\SR(\bC_0, \ldots, \bC_{m-1})$ has dimension  $m(mt-\sum\limits_{i=0}^{m-1}|{\bf T}_i|)$
		and
		$d_{sr}(\SR({\bf C}_0,...,{\bf C}_{m-1}))$ is at least
		\begin{eqnarray*}
		\max\left\{
		\begin{array}{l}
	\min\{m(\delta_0+s_0),(m-1)(\delta_1+s_1),...,\delta_{m-1}+s_{m-1}\}, \\
		\min\{\delta_0+s_0,2(\delta_1+s_1),...,m(\delta_{m-1}+s_{m-1})\}	
		\end{array}
		\right\}
		\end{eqnarray*}		
		\end{theorem}
	
	\begin{proof}
		{\rm The desired conclusions follow from Theorem \ref{thm-mainmain}} and the Hartmann-Tzeng bound for cyclic codes in the Hamming metric.
	\end{proof}
	
	When all ${\bf C}_i$ are BCH cyclic codes, we call the constructed  linear sum-rank code
	$\SR(\bC_0, \ldots, \bC_{m-1})$ an {\it BCH type cyclic sum-rank code},  for distinguishing it from those sum-rank BCH codes in \cite{MP21}.  Many BCH type cyclic sum-rank codes were given in \cite{Chen1}.  \\

 For a given designed sum-rank distance, lower bounds on the dimension of the corresponding cyclic-skew-cyclic sum-rank code are given in sum-rank BCH bound (SR-BCH) and sum-rank Hartmann-Tzeng bound (SR-HT), in \cite{MP21,ALNWZ}. For a given designed sum-rank distance, lower bounds on the dimension of the corresponding cyclic sum-rank code $SR(\bC_0, \ldots, \bC_{m-1})$ are given in our BCH bound in Theorem 3.2 (BCH-CSR) and the Hartmann-Tzeng bound in Theorem 3.3 (HT-CSR). In the following table, we list and compare these lower bounds on dimensions for binary linear sum-rank codes with block lengths $109$, $113$, $137$, $139$, $149$, $157$, $159$, $163$, $173$, $175$, $181$, $225$ and $229,$ and the matrix size $2\times 2$. It is obvious that our bounds for cyclic sum-rank codes are better. \\

 \begin{longtable}{|c|c|c|c|c|c|}
 	\caption{\label{tab:A-q-5-3} $q=2$ and $m=2$}\\ \hline
 	Block length & $d_{sr}\geq$&dim(SR$\cdot$BCH)&dim(BCH$\cdot$CSR)&dim(SR$\cdot$HT)&dim(HT$\cdot$CSR) \\ \hline
 	
 	\multirow{1}*{$t=109$}
 	&$5$& $ 2 \cdot 146$  &$2 \cdot 164$  & $ 2 \cdot 182$ &$2 \cdot 182$ \\ \hline
 	
 	\multirow{1}*{$t=113$}
 	
 	&$5$& $ 2 \cdot 170$  &$2 \cdot 184$  & $ 2 \cdot 198$ &$2 \cdot 198$ \\ \hline
 	
 	\multirow{3}*{$t=137$}
 	
 	&$5$& $ 2 \cdot 138$  &$2 \cdot 172$  & $ 2 \cdot 206$ &$2 \cdot 206$ \\ \cline{2-6}
 	&$6$& $ 2 \cdot 138$  &$2 \cdot 172$  & $ 2 \cdot 204$ &$2 \cdot 205$ \\ \cline{2-6} 		
 	&$9$ & $2 \cdot 70$   &$2 \cdot 104$  & $ 2 \cdot 138$ &$2 \cdot 172$ \\ \hline

 	\multirow{1}*{$t=139$}
 	
 	&$7$& $ 2 \cdot 2$  &$2 \cdot 71$  & $ 2 \cdot 140$ &$2 \cdot 140$  \\\hline	
 	
 	\multirow{1}*{$t=149$}
 	
 	&$8$& $ 2 \cdot 2$   &$2 \cdot 76$  & $ 2 \cdot 150$ &$2 \cdot 150$ \\ \hline
 	
 	\multirow{2}*{$t=157$}
 	
 	&$5$& $ 2 \cdot 210$  &$2 \cdot 236$  & $ 2 \cdot 262$ &$2 \cdot 262$ \\ \cline{2-6}
 	&$8$& $ 2 \cdot 158$  &$2 \cdot 210$  & $ 2 \cdot 210$ &$2 \cdot 236$ \\ \hline
 	
 	\multirow{1}*{$t=159$}
 	
 	&$9$ & $2 \cdot 110$  &$2 \cdot 136$  & $ 2 \cdot 162$ &$2 \cdot 162$ \\ \hline	
 	
 	\multirow{1}*{$t=163$}
 	
 	&$8$& $ 2 \cdot 2$    &$2 \cdot 83$  & $ 2 \cdot 164$ &$2 \cdot 164$ \\ \hline
 	
 	\multirow{1}*{$t=173$}
 	
 	&$8$& $ 2 \cdot 2$  &$2 \cdot 88$  & $ 2 \cdot 174$ &$2 \cdot 174$ \\ \hline
 	
 	\multirow{1}*{$t=175$}
 	&$4$& $ 2 \cdot 284$  &$2 \cdot 316$  & $ 2 \cdot 318$ &$2 \cdot 333$ \\ \hline	
 	
 	\multirow{1}*{$t=181$}
 	
 	&$8$& $ 2 \cdot 2$  &$2 \cdot 92$ & $ 2 \cdot 182$ &$2 \cdot 182$ \\ \hline
 	
 	\multirow{1}*{$t=225$}
 	&$4$& $ 2 \cdot 364$  &$2 \cdot 406$ & $ 2 \cdot 418$  &$2 \cdot 433$ \\ \hline	
 	
 	\multirow{4}*{$t=229$}
 	
 	&$5$& $ 2 \cdot 306$  &$2 \cdot 344$  & $ 2 \cdot 382$ &$2 \cdot 382$ \\ \cline{2-6}	
 	&$6$& $ 2 \cdot 306$  &$2 \cdot 344$  & $ 2 \cdot 380$ &$2 \cdot 381$ \\ \cline{2-6}	
 	&$7$& $ 2 \cdot 230$  &$2 \cdot 306$  & $ 2 \cdot 306$ &$2 \cdot 344$ \\ \cline{2-6}	
 	&$12$ &$2 \cdot 154$  &$2 \cdot 230$  & $ 2 \cdot 230$ &$2 \cdot 305$  \\ \hline	
 		
 \end{longtable}

\section{Specific families of cyclic sum-rank codes}\label{sec-3}

\subsection{Cyclic sum-rank codes from BCH cyclic codes}\label{sec-sub4.1}

Throughout this section,  let $t>1$ be an integer with $\gcd(q,t)=1$ and $m \geq 1$ be an integer. Define
$\ell =\ord_{t}(q^m)$.
We now recall the BCH cyclic codes over $\bF_{q^m}$.
Let $\alpha$ be a primitive element of $\bF_{q^{m\ell}}$ and put $\beta=\alpha^{(q^{m\ell}-1)/t}$.
Then $\beta$ is a $t$-th root of unity.
The {\it minimal polynomial} $\m_{\beta^i}(x)$ of $\beta^i$ over ${\mathbf F}_{q^m}$ is the monic polynomial of the smallest degree over ${\mathbf F}_{q^m}$ with $\beta^i$ as a root.
A Bose-Chaudhuri-Hocquenghem (BCH) code over ${\bf F}_{q^m}$ with code length $t$, designed distance
$\delta$, denoted by ${\bf C}_{(q^m,t,\delta,b)}$, is the cyclic code with generator polynomial
\begin{eqnarray}\label{eqn-original01}
g_{(q^m,t,\delta,b)}(x)={\rm lcm}( \m_{\beta^{b}}(x), \m_{\beta^{b+1}}(x),...,  \m_{\beta^{b+\delta-2}}(x)),
\end{eqnarray}
	 where the least common multiple is computed in ${\mathbf F}_{q^m}[x]$. 	
	 When $b= 1$, the code $\mathbf{C}_{(q^m,t,\delta,b)}$ is called a {\it narrow-sense BCH code}.
	 If $t = q^{m\ell} -1$ (or $q^{m\ell} +1$), then $\mathbf{C}_{(q^m,t,\delta,b)}$ is referred to as a
	 {\it primitive} (or {\it antiprimitive}) BCH code.
It follows from the BCH bound for cyclic codes that $d_H(\bC_{(q^m,t,\delta,b)}) \geq \delta$.   	
For a survey of BCH cyclic codes, the reader is referred to \cite{Ding1}. 	\\

The following
theorem then follows from Theorem \ref{T-2-8}.

\begin{theorem}\label{thm-BCHbridge}
Let $\bC_{(q^m,t,\delta_i,b_i)}$ be a BCH cyclic code of length $t$ and dimension $k_i$ over $\bF_{q^m}$ defined above,
$0 \leq i \leq m-1$.  Then $\SR(\bC_{(q^m,t,\delta_0,b_0)}, \ldots, \bC_{(q^m,t,\delta_{m-1},b_{m-1})})$ is
a cyclic sum-rank code over $\bF_q$ with dimension $m\sum_{i=0}^{m-1}k_i$ and minimum sum-rank
distance at least
		  $$
		   \max\{ \min \{m\delta_0, (m-1)\delta_1,...,\delta_{m-1}\},
		  \min \{\delta_0, 2\delta_1,...,m\delta_{m-1}\}
		   \}
		  $$	
\end{theorem}

	In some cases,  the minimum sum-rank distances of some cyclic sum-rank codes can be determined explicitly.  The following Proposition \ref{prop-20241} demonstrates this possibility.
	
		\begin{lemma}{\citep{Ding4}}\label{L-4-1}
		Let $\ell' =2h$, where $h$ is a positive integer. Let $n=q^{\ell'}-1$.
		For $1\leq \eta\leq q-1,$ the primitive BCH code $C_{(q,n,\eta(q^h+1),1)}$ has parameters$$[q^{\ell'}-1,n-\ell'\left(2\eta(q^h-q^{h-1})-(\eta-1)^2\right)/2,\eta(q^h+1)].$$
	\end{lemma}
	
	\begin{proposition}\label{prop-20241}
		Let $\ell =2h$, where $h$ is a positive integer. Let $t=q^{2\ell}-1$. For $1\leq \eta, 2\eta\leq q-1,$ the dimension of the cyclic sum-rank code $\SR(C_{(q^2,t,\eta(q^{2h}+1),1)},C_{(q^2,t,2\eta(q^{2h}+1),1)})$ is
		$$2\cdot\left[2t-\ell\left( 6\eta(q^{2h}-q^{2(h-1)})-(\eta-1)^2-(2\eta-1)^2\right)\right],$$
		 and the minimum sum-rank distance of
		 $\SR(C_{(q^2,t,\eta(q^{2h}+1),1)},C_{(q^2,t,2\eta(q^{2h}+1),1)})$ is $2\eta(q^{2h}+1).$
	\end{proposition}
	
\noindent
{\bf Proof:} The conclusion follows directly from Proposition \ref{P-3.2} and Lemma \ref{L-4-1}.  \\

BCH cyclic codes in the Hamming metric have wide applications in communication and data storage systems,
as they have very good error-correcting capability and efficient decoding algorithms.  In the past ten years, a lot of
progress on the study of BCH cyclic codes in the Hamming metric has been made.  In many cases, the
dimension  of the BCH code $\bC_{(q^m,t,\delta_i,b_i)}$ is known. In some cases, both the dimension
and minimum distance of $\bC_{(q^m,t,\delta_i,b_i)}$ are known. All known results on the BCH codes $\bC_{(q^m,t,\delta_i,b_i)}$ can be plugged into Theorem \ref{thm-BCHbridge} and a large amount of
results on the cyclic sum-rank codes $\SR(\bC_{(q^m,t,\delta_0,b_0)}, \ldots, \bC_{(q^m,t,\delta_{m-1},b_{m-1})})$ can be  obtained. Known results on  the BCH cyclic codes $\bC_{(q^m,t,\delta_i,b_i)}$ can be found in \cite{ASA,Ding2,Ding3,Ding4,Ding1,Li,Li2,LCJ,LH,YLLY}.

\begin{example}\label{exam-3.1}
{\em
Let $(q, m, t)=(2,2,7)$ and let $(\delta_0, b_0)=(2,0)$ and $(\delta_1, b_1)=(2,1)$.
Then $\SR(\bC_{(q^m,t,\delta_0,b_0)}$ and $ \bC_{(q^m,t,\delta_{1},b_{1})})$
have parameters $[7,6,2]$ and $[7,4,3]$, respectively.
The cyclic sum-rank code $\SR(\bC_{(q^m,t,\delta_0,b_0)}, \bC_{(q^m,t,\delta_{1},b_{1})})$
 has block length $7$, dimension $20$, and
minimum sum-rank distance $3$.
}
\end{example}

\begin{example}\label{exam-3.2}
{\em
Let $(q, m, t)=(2,2,9)$ and let $(\delta_0, b_0)=(3,0)$ and $(\delta_1, b_1)=(2,0)$.
Then $\SR(\bC_{(q^m,t,\delta_0,b_0)}$ and $ \bC_{(q^m,t,\delta_{1},b_{1})})$
have parameters $[9,5,3]$ and $[9,8,2]$, respectively.
The cyclic sum-rank code $\SR(\bC_{(q^m,t,\delta_0,b_0)}, \bC_{(q^m,t,\delta_{1},b_{1})})$
 has block length $9$, dimension $26$, and
minimum sum-rank distance at least $3$.
}
\end{example}

\subsection{Cyclic sum-rank codes from other cyclic codes}

There are a number of known families of cyclic codes over finite fields in the Hamming metric whose dimensions are known and whose minimum distances are known or lower bounded. Any multiset of them
can be plugged into Theorem \ref{thm-mainmain} and a cyclic sum-rank code
$\SR(\bC_0, \ldots, \bC_{m-1})$ with known dimension and a lower bound on its minimum sum-rank
distance is obtained.   The following is a short list of them:
\begin{itemize}
\item Certain families of irreducible cyclic codes.
\item Certain families of BCH cyclic codes.
\item The punctured Reed-Muller codes
\item The punctured Dilix codes.
\item Families of cyclic codes with a few weights.
\end{itemize}
These are certainly new and specific families of cyclic sum-rank codes with known dimensions and controllable minimum sum-rank distances.  Below are four example of such codes $\SR(\bC_0, \bC_{1})$.

\begin{example}\label{exam-4.1}
{\em
Let $(q, m, t)=(3,2,5)$. Let $\alpha$ be a generator of $\bF_{q^2}^*$ with $\alpha^2+2\alpha+2=0.$
Let $\bC_0$ be the cyclic code of length $t$ over $\bF_{q^2}$ with generator polynomial
$x^2 + \alpha x + 1$.
Then $\bC_0$ has parameters $[5, 3, 3]$.
Let $\bC_1$ be the cyclic code of length $t$ over $\bF_{q^2}$ with generator polynomial
$x^2 + \alpha^3 x + 1$.
Then $\bC_1$ has parameters $[5, 3, 3]$.
The cyclic sum-rank code $\SR(\bC_0, \bC_1)$ has block length $5$, dimension $12$, and
minimum sum-rank distance $3$.
Let $d_i=d_H(\bC_i)$. Then $\min\{2d_0, d_1\} = \min\{d_0,2 d_1\}=3$. Hence,  the lower bound
on the minimum sum-rank distance of $\SR(\bC_0, \bC_1)$ documented in Theorem \ref{thm-mainmain}
is achieved.
}
\end{example}

\begin{example}\label{exam-4.2}
{\em
Let $(q, m, t)=(3,2,5)$. Let $\alpha$ be a generator of $\bF_{q^2}^*$ with $\alpha^2+2\alpha+2=0.$
Let $\bC_0$ be the cyclic code of length $t$ over $\bF_{q^2}$ with generator polynomial
$(x^2 + \alpha x + 1)(x-1)$.
Then $\bC_0$ has parameters $[5, 2, 4]$.
Let $\bC_1$ be the cyclic code of length $t$ over $\bF_{q^2}$ with generator polynomial
$x^2 + \alpha^3 x + 1$.
Then $\bC_1$ has parameters $[5, 3,3]$.
The cyclic sum-rank code $\SR(\bC_0, \bC_1)$ has block length $5$, dimension $10$, and
minimum sum-rank distance $4$.
Let $d_i=d_H(\bC_i)$. Then $\min\{2d_0, d_1\} <  \min\{d_0,2 d_1\}=4$. Hence,  the lower bound
on the minimum sum-rank distance of $\SR(\bC_0, \bC_1)$ documented in Theorem \ref{thm-mainmain}
is achieved.
}
\end{example}

\begin{example}\label{exam-4.3}
{\em
Let $(q, m, t)=(3,2,5)$. Let $\alpha$ be a generator of $\bF_{q^2}^*$ with $\alpha^2+2\alpha+2=0.$
Let $\bC_0$ be the cyclic code of length $t$ over $\bF_{q^2}$ with generator polynomial
$x^2 + \alpha x + 1$.
Then $\bC_0$ has parameters $[5, 3, 3]$.
Let $\bC_1$ be the cyclic code of length $t$ over $\bF_{q^2}$ with generator polynomial
$(x^2 + \alpha^3 x + 1)(x-1)$.
Then $\bC_1$ has parameters $[5, 2, 4]$.
The cyclic sum-rank code $\SR(\bC_0, \bC_1)$ has block length $5$, dimension $10$, and
minimum sum-rank distance $4$.
Let $d_i=d_H(\bC_i)$. Then $4=\min\{2d_0, d_1\} > \min\{d_0,2 d_1\}$. Hence,  the lower bound
on the minimum sum-rank distance of $\SR(\bC_0, \bC_1)$ documented in Theorem \ref{thm-mainmain}
is achieved.
}
\end{example}

\begin{example}\label{exam-4.4}
{\em
Let $(q, m, t)=(3,2,5)$. Let $\alpha$ be a generator of $\bF_{q^2}^*$ with $\alpha^2+2\alpha+2=0.$
Let $\bC_0$ be the cyclic code of length $t$ over $\bF_{q^2}$ with generator polynomial
$x-1$.
Then $\bC_0$ has parameters $[5, 4, 2]$.
Let $\bC_1$ be the cyclic code of length $t$ over $\bF_{q^2}$ with generator polynomial
$(x^2 + \alpha^3 x + 1)(x-1)$.
Then $\bC_1$ has parameters $[5, 2,4]$.
The cyclic sum-rank code $\SR(\bC_0, \bC_1)$ has block length $5$, dimension $12$, and
minimum sum-rank distance $4$.
Let $d_i=d_H(\bC_i)$. Then $4=\min\{2d_0, d_1\} >  \min\{d_0,2 d_1\}$. Hence,  the lower bound
on the minimum sum-rank distance of $\SR(\bC_0, \bC_1)$ documented in Theorem \ref{thm-mainmain}
is achieved.   \\

Note that the Singleton-like bound asserts that
$$
\dim(\bC) \leq m(mt -d_{sr}+1)
$$
for any linear sum-rank code with block length $t$,  matrix size $m \times m$. and minimum sum-rank
distance $d_{sr}$.
Since the Singleton-like bound above is a multiple of $m$,  this ternary cyclic sum-rank code $\SR(\bC_0, \bC_1)$ is almost-optimal with respect to the Singleton-like bound.  The authors are not aware of any
ternary linear sum-rank code of block length $5$ and dimension $12$ that has a larger minimum sum-rank
distance.
}
\end{example}

\section{Negacyclic sum-rank codes}\label{sec-4}
	
There are a number of known families of negacyclic codes over finite fields in the Hamming metric whose dimensions are known and whose minimum distances are known or lower bounded (\cite{SDW23,WDLZ,WX,WTD23,ZKZL,ZH}).   Any multiset of them
can be plugged into Theorem \ref{thm-mainmain} and a negacyclic sum-rank code
$\SR(\bC_0, \ldots, \bC_{m-1})$ with known dimension and a lower bound on its minimum sum-rank
distance is obtained.   As an example, below we use the family of BCH negacyclic codes to illustrate this technique.  \\

 Let $q$ be a power of an odd prime, and let $t$ be positive integer with
 $\gcd(t, q)=1$ and $m$ be a positive integer.
Let $\alpha$ be a primitive element of $\bF_{q^{m\ell}}$, where $\ell=\ord_{2t}(q^m)$ is the order of $q^m$ modulo $2t$. Put $\beta=\alpha^{(q^{m\ell}-1)/2t}$,
then $\beta$ is a primitive $2t$-th root of unity in $\bF_{q^{m\ell}}$.
 For any $i$ with $0\leq i\leq n-1$,
let $\m_{\beta^{1+2i}}(x)$ denote the minimal polynomial of $\beta^{1+2i}$ over $\bF_{q^m}$.  For any $\delta$
 with $2 \leq \delta \leq t$, let
\begin{eqnarray}\label{eqn-main2}
g_{(q^m,t,\delta,b)}(x)=\lcm \left(\m_{\beta^{1+2b}}(x), \m_{\beta^{1+2(b+1)}}(x), \ldots, \m_{\beta^{1+2(b+\delta-2)}}(x)\right),
\end{eqnarray}
where $b$ is an integer and lcm denotes the least common multiple of these minimal polynomials.
Let $\bC_{(q^m,t,\delta,b)}$ denote
the negacyclic code of length $t$ over $\bF_{q^m}$ with generator polynomial $g_{(q^m,t,\delta,b)}(x)$, then $\bC_{(q^m,t,\delta,b)}$ is called a \emph{BCH negacyclic code} with {\it designed distance} $\delta$. It is known that $d_H(\bC_{(q^m,t,\delta,b)}) \geq \delta.$ \\

The following theorem follows from Theorem \ref{thm-mainmain} and the fact that $d_H(\bC_{(q^m,t,\delta,b)}) \geq \delta.$

\begin{theorem}\label{thm-BCHnegabridge}
Let $\bC_{(q^m,t,\delta_i,b_i)}$ be a BCH negacyclic code of length $t$ and dimension $k_i$ over $\bF_{q^m}$ defined above,
$0 \leq i \leq m-1$.  Then $\SR(\bC_{(q^m,t,\delta_0,b_0)}, \ldots, \bC_{(q^m,t,\delta_{m-1},b_{m-1})})$ is
a negacyclic sum-rank code over $\bF_q$ with block size $t$, matrix size $m \times m$,  dimension $m\sum_{i=0}^{m-1}k_i$ and minimum sum-rank
distance at least
		  $$
		   \max\{ \min \{m\delta_0, (m-1)\delta_1,...,\delta_{m-1}\},
		  \min \{\delta_0, 2\delta_1,...,m\delta_{m-1}\}
		   \}
		  $$	
\end{theorem}

In some cases, the dimension of $\bC_{(q^m,t,\delta_i,b_i)}$ is known \cite{WX}. When such a multiset
of $m$ negacyclic codes $\bC_{(q^m,t,\delta_i,b_i)}$ are plugged into Theorem \ref{thm-BCHnegabridge},
a ngacyclic sum-rank code  $\SR(\bC_{(q^m,t,\delta_0,b_0)}, \ldots, \bC_{(q^m,t,\delta_{m-1},b_{m-1})})$ with known dimension and a lower bound on
its minimum sum-rank distance is obtained.

\begin{example}\label{exam-5.1}
{\em
Let $(q, m, t)=(3,2,5)$. Let $\alpha$ be a generator of $\bF_{q^2}^*$ with $\alpha^2+2\alpha+2=0.$
The canonical factorisation of $x^t+1$ over $\bF_{q^2}$ is
$$
x^5+1=(x+1)(x^2 + \alpha^5 x + 1)(x^2 + \alpha^7 x + 1).
$$
Let $\bC_0$ be the negacyclic code of length $t$ over $\bF_{q^2}$ with generator polynomial
$x^2 + \alpha^5 x + 1$.
Then $\bC_0$ has parameters $[5, 3, 3]$.
Let $\bC_1$ be the negacyclic code of length $t$ over $\bF_{q^2}$ with generator polynomial
$x^2 + \alpha^7 x + 1$.
Then $\bC_1$ has parameters $[5, 3, 3]$.
The negacyclic sum-rank code $\SR(\bC_0, \bC_1)$ has block length $5$, dimension $12$, and
minimum sum-rank distance $3$.
Let $d_i=d_H(\bC_i)$. Then $\min\{2d_0, d_1\} = \min\{d_0,2 d_1\}=3$. Hence,  the lower bound
on the minimum sum-rank distance of $\SR(\bC_0, \bC_1)$ documented in Theorem \ref{thm-mainmain}
is achieved.
}
\end{example}

\begin{example}\label{exam-5.2}
{\em
Let $(q, m, t)=(3,2,5)$. Let $\alpha$ be a generator of $\bF_{q^2}^*$ with $\alpha^2+2\alpha+2=0.$
The canonical factorisation of $x^t+1$ over $\bF_{q^2}$ is
$$
x^5+1=(x+1)(x^2 + \alpha^5 x + 1)(x^2 + \alpha^7 x + 1).
$$
Let $\bC_0$ be the negacyclic code of length $t$ over $\bF_{q^2}$ with generator polynomial
$x + 1$.
Then $\bC_0$ has parameters $[5, 4, 2]$.
Let $\bC_1$ be the negacyclic code of length $t$ over $\bF_{q^2}$ with generator polynomial
$(x^2 + \alpha^7 x + 1)(x+1)$.
Then $\bC_1$ has parameters $[5, 2, 4]$.
The negacyclic sum-rank code $\SR(\bC_0, \bC_1)$ has block length $5$, dimension $12$, and
minimum sum-rank distance $4$.
Let $d_i=d_H(\bC_i)$. Then $4=\min\{2d_0, d_1\} > \min\{d_0,2 d_1\}$. Hence,  the lower bound
on the minimum sum-rank distance of $\SR(\bC_0, \bC_1)$ documented in Theorem \ref{thm-mainmain}
is achieved.  This code is better than the code in Example \ref{exam-5.1}.
}
\end{example}

\section{Constacyclic sum-rank codes}\label{sec-n1}

Similarly,  $\lambda$-constacyclic sum-rank codes are defined as follows.  They are a generalization of
the cyclic and negacyclic sum-rank codes.

\begin{definition}[Constacyclic sum-rank code]
		Let $\lambda$ be a nonzero element in $\bF_{q}$.  A sum-rank code ${\bf C}$ over ${\bf F}_q$ with block length $t$ and matrix size $n \times m$ is called a {\it $\lambda$-constacyclic sum-rank code}, if ${\bf x}=({\bf x}_1,\ldots,{\bf x}_t)$ is a codeword in ${\bf C}$,
		${\bf x}_i \in {\bf F}_q^{(n,m)}$, $i=1,\ldots,t$,  implies that
		$(\lambda {\bf x}_{t},{\bf x}_1,...,{\bf x}_{t-1})$ is also a codeword in ${\bf C}$.
\end{definition}

The following theorem follows directly from Theorem \ref{thm-mainmain}.

\begin{theorem}\label{thm-mainmain3}
Let $\lambda$ be a nonzero element in $\bF_{q}$.
	Let ${\bf C}_i \subset {\bf F}_{q^m}^t$ be a $[t, k_i, d_i]_{q^m}$ $\lambda$-constacyclic code
	over ${\bf F}_{q^m}$, $0 \leq i \leq m-1$. Then $\SR({\bf C}_0,...,{\bf C}_{m-1})$ is a
	$\lambda$-constacyclic sum-rank code  over ${\bf F}_q$ with block length $t$, matrix size $m \times m$ and dimension  $m(k_0+\cdots+k_{m-1})$.
	The minimum sum-rank distance of $\SR({\bf C}_0,{\bf C}_{1},...,{\bf C}_{m-1})$ is at least
	$$\max\{\min \{md_0, (m-1)d_1,...,d_{m-1}\}, \min\{d_0,2d_1, \ldots, md_{m-1}\}\}.$$
\end{theorem}

Similarly,  any multiset of $m$ $\lambda$-constacyclic codes of block length $t$ over $\bF_{q^m}$
with known dimensions and controllable minimum Hamming distances can be
plugged into Theorem \ref{thm-mainmain3} and a $\lambda$-constacyclic sum-rank code  over ${\bf F}_q$
with known dimension and controllable minimum sum-rank distance is obtained.  A number of families of
$\lambda$-constacyclic codes over $\bF_{q^m}$
with known dimensions and controllable minimum Hamming distances are available in the literature.

\begin{example}\label{exam-n.1}
{\em
Let $(q, m, t)=(7,2,4)$. Let $\alpha$ be a generator of $\bF_{q^2}^*$ with $\alpha^2+6\alpha+3=0.$
The canonical factorisation of $x^t-2$ over $\bF_{q^2}$ is
$$
x^4-2=(x+2)(x+5)(x+\alpha^4)(x+\alpha^{28}).
$$
Let $\bC_0$ be the $2$-constacyclic code of length $t$ over $\bF_{q^2}$ with generator polynomial
$(x+2)(x+\alpha^4)$.
Then $\bC_0$ has parameters $[4, 2, 3]$.
Let $\bC_1$ be the $2$-constacyclic code of length $t$ over $\bF_{q^2}$ with generator polynomial
$(x+5)(x+\alpha^{28})$.
Then $\bC_1$ has parameters $[4, 2, 3]$.
The $2$-constacyclic sum-rank code $\SR(\bC_0, \bC_1)$ has block length $4$, dimension $8$, and
minimum sum-rank distance $3$.
Let $d_i=d_H(\bC_i)$. Then $\min\{2d_0, d_1\} = \min\{d_0,2 d_1\}=3$. Hence,  the lower bound
on the minimum sum-rank distance of $\SR(\bC_0, \bC_1)$ documented in Theorem \ref{thm-mainmain3}
is achieved.
}
\end{example}

	\begin{example}\label{exam-6.2}
		{\em
			Let $(q, m, t)=(4,2,5)$. Let $\alpha$ be a generator of $\bF_{q^2}^*$ with $\alpha^4+\alpha+1=0.$ It is clear that $\alpha^5\in \bF_{q}.$
			The canonical factorisation of $x^t-\alpha^5$ over $\bF_{q^2}$ is
			$$
			x^5-\alpha^5=(x + \alpha)(x + \alpha^4)(x + \alpha^7)(x + \alpha^{10})(x + \alpha^{13}).
			$$
			Let $\bC_0$ be the $\alpha^5$-constacyclic code of length $t$ over $\bF_{q^2}$ with generator polynomial
			$x + \alpha$.
			Then $\bC_0$ has parameters $[5, 4, 2]$.
			Let $\bC_1$ be the $\alpha^5$-constacyclic code of length $t$ over $\bF_{q^2}$ with generator polynomial
			$(x + \alpha)(x + \alpha^4)(x + \alpha^7)$.
			Then $\bC_1$ has parameters $[5, 2, 4]$.
			The $\alpha^5$-constacyclic sum-rank code $\SR(\bC_0, \bC_1)$ has block length $5$, dimension $12$, and
			minimum sum-rank distance $4$.
			Let $d_i=d_H(\bC_i)$. Then $4=\min\{2d_0, d_1\} > \min\{d_0,2 d_1\}$. Hence,  the lower bound
			on the minimum sum-rank distance of $\SR(\bC_0, \bC_1)$ documented in Theorem \ref{thm-mainmain3}
			is achieved.
		}
	\end{example} 	
	
	\begin{example}\label{exam-6.3}
		{\em
			Let $(q, m, t)=(4,2,5)$. Let $\alpha$ be a generator of $\bF_{q^2}^*$ with $\alpha^4+\alpha+1=0.$ It is clear that $\alpha^{10}\in \bF_{q}.$
			The canonical factorisation of $x^t-\alpha^{10}$ over $\bF_{q^2}$ is
			$$
			x^5-\alpha^{10}=(x + \alpha^2)(x + \alpha^5)(x + \alpha^8)(x + \alpha^{11})(x + \alpha^{14}).
			$$
			Let $\bC_0$ be the $\alpha^{10}$-constacyclic code of length $t$ over $\bF_{q^2}$ with generator polynomial
			$x + \alpha^2$.
			Then $\bC_0$ has parameters $[5, 4, 2]$.
			Let $\bC_1$ be the $\alpha^{10}$-constacyclic code of length $t$ over $\bF_{q^2}$ with generator polynomial
			$(x + \alpha^2)(x + \alpha^5)(x + \alpha^8)$.
			Then $\bC_1$ has parameters $[5, 2, 4]$.
			The $\alpha^{10}$-constacyclic sum-rank code $\SR(\bC_0, \bC_1)$ has block length $5$, dimension $12$, and minimum sum-rank distance $4$.
			Let $d_i=d_H(\bC_i)$. Then $4=\min\{2d_0, d_1\} > \min\{d_0,2 d_1\}$. Hence,  the lower bound
			on the minimum sum-rank distance of $\SR(\bC_0, \bC_1)$ documented in Theorem \ref{thm-mainmain3}
			is achieved.
		}
	\end{example}

\section{Distance-optimal binary cyclic sum-rank codes}\label{sec-chenh}

In this section, we construct infinitely many distance-optimal cyclic sum-rank codes in ${\bf F}_2^{(2,2), \ldots,(2,2)}$ ($t$ copies) with the minimum sum-rank distance four. First of all, let $V_{sr}(2)$ be the number of elements in the ball with the radius $2$ in the sum-rank metric space ${\bf F}_2^{(2,2), \ldots,(2,2)}$. Notice that there is only one rank-zero $2 \times 2$ matrix over ${\bf F}_2$. There are $9$ rank-one $2 \times 2$ matrices over ${\bf F}_2$ and $6$ rank-two $2 \times 2$ matrices over ${\bf F}_2$. Then we have $$V_{sr}(2)=1+15t+81 \cdot \frac{t(t-1)}{2}.$$

\begin{theorem}\label{thm-c7.1}
If $\frac{81t(t-1)}{2} \geq 2^r$, a codimension-$r$ binary linear sum-rank code with block length $t$,  matrix size $2 \times 2$, and minimum sum-rank distance $4$,  is distance-optimal with respect to the sphere packing bound in the sum-rank metric.
\end{theorem}

\begin{proof}
{\em
From the condition specified in this theorem,  we have
$$V_{sr}(2) > 2^r.$$
It then follows from the sphere packing bound in the sum-rank metric \cite{BGR} that there is no block length $t$ and codimension $r$ binary sum-rank code with minimum distance $5$. The conclusion is proved.
}
\end{proof}

Throughout this section,  let $\bC_0$ be the cyclic code of length $t$ over $\bF_4$ with generator
polynomial $x-1$.  It is easily seen that $\bC_0$ has parameters $[t, t-1, 2]_4$ and is MDS.

\begin{corollary}\label{cor-c7.1}
If there is a linear quaternary $[t, k_1, 4]_4$ code $\bC_1$ satisfying $$\frac{81t(t-1)}{2} \geq 4^{t-k_1+1},$$ then $SR(\bC_0,\bC_1)$ is a binary linear distance-optimal sum-rank code.
\end{corollary}

\begin{proof}
{\em
The minimum sum-rank distance of $SR(\bC_0, \bC_1)$ is at least $4$ from Theorem \ref{thm-mainmain}. The dimension of $\SR(\bC_0,\bC_1)$ over ${\bf F}_2$ is $2(t-1+k_1)$, the codimension is $2(t-k_1)+2$. The desired conclusion follows immediately.
}
\end{proof}

The following result follows from Corollary \ref{cor-c7.1}  immediately.

\begin{corollary}\label{cor-c7.2}
Let $t$ be a positive integer satisfying $t \geq 12$. If there is a linear quaternary $[t, t-5, 4]_4$ code
$\bC_1$,
then $\SR(\bC_0,\bC_1)$ is  a distance-optimal binary sum-rank code.
\end{corollary}

Using those optimal linear quaternary codes with minimum distance $4$ in \cite{codetable} as the code
$\bC_1$ in Corollary \ref{cor-c7.2},  many distance-optimal binary sum-rank codes with  minimum sum-rank distance $4$ are obtained.  For example, for block lengths $t$ satisfying $12 \leq t \leq 23$, we obtain distance-optimal binary sum-rank codes $\SR(\bC_0,\bC_1)$ with minimum sum-rank distance $4$ from those codes ${\bf C}_1$ in \cite{codetable}.\\

We have the following infinite family of distance-optimal binary cyclic sum-rank codes.

\begin{corollary}\label{cor-c7-3}
Let $t=4^\ell-1$, where $\ell \geq 2$.  Let $\bC_1$ denote the BCH cyclic code $\bC_{(4, t, 4, 0)}$ defined
in Section \ref{sec-sub4.1}.  Then $\SR(\bC_0, \bC_1)$ has  block length $4^\ell-1$, matrix size $2 \times 2$,
and dimension $4(4^\ell-\ell-2)$ and minimum sum-rank distance $4$. Furthermore,
the binary sum-rank cyclic code $\SR(\bC_0, \bC_1)$ is distance-optimal.
\end{corollary}

\begin{proof}
{\em
It follows from the BCH bound on cyclic codes that the minimum distance of $\bC_1$ is at least $4$.
By the sphere-packing bound, the minimum distance of $\bC_1$ is at most $4$.  Hence,  $\bC_1$ has
minimum distance $4$.  It is well known that $\bC_1$ has dimension $4^\ell -2\ell -2$.
The desired parameters of the binary sum-rank code $\SR(\bC_0, \bC_1)$ then follow from Proposition \ref{P-3.2}.   Since both $\bC_0$ and $\bC_1$ are cyclic, $\SR(\bC_0, \bC_1)$ is cyclic.
It is easy to verify that $$\frac{81(4^\ell-1)(4^\ell-2)}{2} \geq 4^{2\ell+2}.$$
It then follows from Corollary \ref{cor-c7.1} that $\SR(\bC_0, \bC_1)$ is distance-optimal.
}
\end{proof}

\section{Decoding of the cyclic or negacyclic sum-rank codes $\SR({\bf C}_0, \ldots, {\bf C}_{m-1})$}\label{sec-5}

The decoding of a binary sum-rank code $\SR({\bf C}_0, {\bf C}_1)$ of matrix size $2\times 2$ can be
reduced to the decoding of the two quaternary codes ${\bf C}_0$ and ${\bf C}_1$,
if the conditions $d_H({\bf C}_0) \geq d_{sr}$ and $d_H({\bf C}_1) \geq \frac{2d_{sr}}{3}$ on two quaternary cyclic codes ${\bf C}_0$ and ${\bf C}_1$ are satisfied \citep{Chen1}.  Similarly, the decoding of a
$q$-ary cyclic or negacyclic sum-rank code $\SR({\bf C}_0, \ldots, {\bf C}_{m-1})$ of matrix size
$m \times m$ can be reduced to the decoding of $m$ cyclic or negacyclic codes $\bC_i$ over
$\bF_{q^m}$ under certain conditions by extending the decoding techniques in \cite{Chen1,ChenReed}.

\section{Concluding remarks}\label{sec-6}

It is very hard to determine the dimension of a linear sum-rank code with block size
$t \geq 2$ and it is much harder to settle its minimum sum-rank distance.  In this situation,
it is interesting to construct an infinite family of linear sum-rank codes with known dimensions
and reasonable lower bounds on their minimum sum-rank distances.  \\

Cyclic, negacyclic and constacyclic sum-rank codes of the type $\SR(\bC_0, \ldots, \bC_{m-1})$ were introduced
in this paper.
The cyclic-skew-cyclic sum-rank codes and sum-rank BCH codes introduced and constructed in \cite{MP21,ALNWZ} are special cases of the cyclic sum-rank codes introduced in this paper.
Specific constructions of cyclic, negacyclic and constacyclic sum-rank codes of the type $\SR(\bC_0, \ldots, \bC_{m-1})$  directly from cyclic, negacyclic and constacyclic codes in the Hamming metric were presented in this paper.
When the dimensions of the underlying  cyclic, negacyclic and constacyclic codes in the Hamming metric are known
and their minimum Hamming distances have a good lower bound, the corresponding  cyclic, negacyclic
and constacyclic sum-rank codes $\SR(\bC_0, \ldots, \bC_{m-1})$ have a known dimension and lower bound on its
minimum sum-rank distance.   In some cases, it is not hard to determine minimum sum-rank distances of cyclic, negacyclic and constacyclic sum-rank codes of the type $\SR(\bC_0, \ldots, \bC_{m-1})$. The main bridge used in this paper is Theorem \ref{thm-mainmain}. It would be a breakthrough if the lower bound on the minimum sum-rank distance of the code
$\SR(\bC_0, \ldots, \bC_{m-1})$ documented in Theorem \ref{thm-mainmain} can be improved.  \\

As one application of our construction of cyclic sum-rank codes, an infinite family of distance-optimal binary cyclic sum-rank codes were given in this paper. This is the first infinite family of optimal sum-rank codes with minimum sum-rank distance four. It is a challenging problem to construct distance-optimal sum-rank codes with bigger minimum sum-rank distances.\\

\end{document}